\documentclass[aps,prd,superscriptaddress,amsmath,%
                twocolumn,showpacs,10pt]{revtex4-1}
\usepackage{amssymb,graphicx}
\usepackage{dcolumn}

\usepackage{times} \usepackage{mathptmx}

\newcommand{\eq}[1]{Eq.\:(\ref{#1})}

\begin{document}
\title{Fast Fits for Lattice QCD Correlators}
\author{K.~Hornbostel}
\affiliation{Southern Methodist University, Dallas, Texas 75275, USA}
\author{G.~P.~Lepage}
\email[]{g.p.lepage@cornell.edu}
\affiliation{Laboratory of Elementary-Particle Physics, Cornell 
University, Ithaca, New York 14853, USA}
\author{C.~T.~H.~Davies}
\author{R.~J.~Dowdall}
\affiliation{SUPA, School of Physics and Astronomy, University of 
Glasgow, Glasgow, G12 8QQ, UK}
\author{H.~Na}
\author{J.~Shigemitsu}
\affiliation{Department of Physics, The Ohio State University, 
Columbus, OH 43210, USA}
\collaboration{HPQCD collaboration}
\homepage{http://www.physics.gla.ac.uk/HPQCD}
\noaffiliation
\date{4 November 2011}

\begin{abstract}
   We illustrate a technique for fitting lattice QCD correlators to 
sums of exponentials that is significantly faster than traditional 
fitting methods\,---\,10--40~times faster for the realistic examples we 
present. Our  examples are drawn from a recent analysis of the 
$\Upsilon$~spectrum, and another recent analysis of the $D\!\to\!\pi$ 
semileptonic form factor. For single correlators, we show how to 
simplify traditional effective-mass analyses.
\end{abstract}

\maketitle
Most physics results in lattice QCD come from fits of lattice 
correlators to sums of exponentials. For example, we study a particular 
hadron by computing Monte Carlo simulation 
estimates~$G_{ab}^\mathrm{MC}(t)$ of hadronic correlators,
\begin{equation} \label{eq-Gab}
   \sum_\mathbf{x}
   \langle0|\Gamma_b(\mathbf{x},t)\Gamma_a(0,0)|0\rangle ,
\end{equation}
with different sources~$\Gamma_a$ and sinks~$\Gamma_b$ that create and 
destroy the hadron. The sum over all spatial sites~$\mathbf{x}$ 
restricts the hadrons to states with zero total three-momentum. Such a 
correlator can be decomposed into contributions from energy 
eigenstates~$|E_j\rangle$ in QCD~\cite{bib-sumexp}:
\begin{equation} \label{eq-fitfcn}
   G_{ab}(t;N) = \sum_{j=1}^N a_j b_j \exp(-E_jt)
\end{equation}
where $E_j$ is the energy, with $E_j\!\ge\!E_{j-1}$, and the amplitudes 
are
matrix elements, with
\begin{align}
   a_j^* &= \langle 0 | \Gamma_a(0,0) | E_j \rangle ,\nonumber \\
   b_j &= \langle 0 | \Gamma_b(0,0) | E_j \rangle .
\end{align}
The physics is in the energies and the matrix elements, and these are
determined by fitting fomula~(\ref{eq-fitfcn}) to the Monte Carlo 
data~$G_{ab}^\mathrm{MC}(t)$ for a variety of sources and sinks.

In principle, the number of terms~$N$ in \eq{eq-fitfcn} is infinite, 
but, in practice, we need only retain a finite number of terms because 
the exponentials suppress high-energy states. The number needed depends 
upon the precision of the simulation data~$G_{ab}^\mathrm{MC}$, but it 
is not uncommon to require $N\!=\!10$ or more terms for good fits to 
accurate data. The fitting process becomes both cumbersome and time 
consuming if many correlators must be fit simultaneously while using 
such large~$N$s. In this paper we introduce a method that can 
dramatically simplify and accelerate such fits.

The key to this new approach lies in how priors are introduced. Two 
types of input data are required for these fits. The first is 
simulation data for the correlators, consisting of Monte Carlo 
averages~$\overline{G}$ for each $a$, $b$ and $t$, and a covariance 
matrix $\sigma^2$ that specifies both the statistical uncertainties in 
each average and the correlations between them:
\begin{equation}
   G_{ab}^\mathrm{MC}(t)\leftrightarrow\left\{\overline{G}_{ab}(t),
   \sigma^2_{ab,a^\prime b^\prime}(t,t^\prime)\right\}
\end{equation}
This data contributes
\begin{align}
   \chi^2_\mathrm{MC}(a_j,&b_j,E_j) =
 \sum_{t,a,b}\sum_{t^\prime,a^\prime,b^\prime}
   \left(G_{ab}(t;N)-\overline{G}_{ab}(t)\right) \nonumber \\
   &\sigma^{-2}_{ab,a^\prime b^\prime}(t,t^\prime)
   \left(G_{a^\prime b^\prime}(t^\prime;N)-
      \overline{G}_{a^\prime b^\prime}(t^\prime)\right)
\end{align}
to the $\chi^2$~function that is minimized by varying parameters~$a_j$, 
$b_j$, and~$E_j$ in a conventional fit.

The second type of input data consists of Bayesian priors for each fit 
parameter. Complicated multi-correlator, multi-parameter fits are 
impossible without \emph{a priori} estimates for each fit 
parameter~\cite{Lepage:2001ym}:
\begin{align}
   a^\mathrm{pr}_j &\equiv \overline{a}_j \pm \sigma_{a_j}, \nonumber 
\\
   b^\mathrm{pr}_j &\equiv \overline{b}_j \pm \sigma_{b_j},  \nonumber 
\\
   E^\mathrm{pr}_j &\equiv \overline{E}_j \pm \sigma_{E_j}.
   \label{eq-prior}
\end{align}
This information is included in a conventional fit by adding extra 
terms to $\chi^2(a_j,b_j,E_j)$:
$\chi^2\!=\!\chi^2_\mathrm{MC}+\chi^2_\mathrm{pr}$ where
\begin{align}
   \chi^2_\mathrm{pr}&(a_j,b_j,E_j) =\nonumber\\ &\sum_{j=1}^N\left\{
   \frac{(a_j-\overline{a}_j)^2}{\sigma_{a_j}^2} +
   \frac{(b_j-\overline{b}_j)^2}{\sigma_{b_j}^2} +
   \frac{(E_j-\overline{E}_j)^2}{\sigma_{E_j}^2}
   \right\} .
\end{align}
The priors can also be combined to give \emph{a priori} estimates for 
the correlators,
\begin{equation}
   G_{ab}^\mathrm{pr}(t;N) \equiv
   \sum_{j=1}^N a_j^\mathrm{pr} b_j^\mathrm{pr} 
\exp(-E_j^\mathrm{pr}t),
\end{equation}
where the means and covariance matrix for $G_{ab}^\mathrm{pr}(t)$ are 
computed, using standard error propagation, from the means and 
covariance matrix of the priors (\eq{eq-prior}).

The cost of a traditional analysis goes up rapidly with the number of 
parameters needed to obtain a good fit. In practice, however, we are 
rarely interested in parameters from the large-$j$ terms in fit 
function~(\ref{eq-fitfcn}), even when these terms are needed for a good 
fit. Rather than including them in the fit, we can incorporate the 
large-$j$ terms into the Monte Carlo data \emph{before} fitting. To do 
this, we use the priors to generate an \emph{a priori} estimate for 
these terms, and then subtract that estimate from the Monte Carlo data. 
This effectively removes the large-$j$ terms from the data. Finally we 
fit the modified data with a simpler formula that includes only 
small-$j$ terms.

More explicitly, we can remove terms having $n\!<\!j\!\le\!N$ by 
replacing $G^\mathrm{MC}_{ab}(t)$ with (first definition)
\begin{equation}\label{eq-altred}
   {\tilde{G}}_{ab}^{\mathrm{MC}}(t;n) \equiv G_{ab}^\mathrm{MC}(t) - 
\Delta G_{ab}^\mathrm{pr}(t;n),
\end{equation}
where
\begin{align}
   \Delta G^\mathrm{pr}_{ab}(t;n) &\equiv
    G_{ab}^\mathrm{pr}(t;N)-
    G_{ab}^\mathrm{pr}(t;n) \nonumber \\
    &=
    \sum_{j=n+1}^N a_j^\mathrm{pr} b_j^\mathrm{pr} 
\exp(-E_j^\mathrm{pr}t)
\end{align}
is the $j\!>\!n$ part of the fit function. Having removed the 
$j\!>\!n$~terms, we can fit~$\tilde G_{ab}^\mathrm{MC}(t;n)$ with the 
simpler fit function,~$G_{ab}(t;n)$, rather than $G_{ab}(t;N)$.

Here we assume that~$N$ is sufficiently large that $\Delta 
G_{ab}^\mathrm{pr}(t;n)$ and therefore 
$\tilde{G}_{ab}^{\mathrm{MC}}(t;n)$ are independent of $N$ to within 
their statistical errors. The covariance matrix for 
$\tilde{G}_{ab}^{\mathrm{MC}}(t;n)$ is obtained by adding the 
covariance matrices of $G_{ab}^{\mathrm{MC}}(t)$ and $\Delta 
G_{ab}^\mathrm{pr}(t;n)$ (that is, adding the errors in 
quadrature)~\cite{bib-errors}.

Removing high-$j$ terms from both the fit function and the fit data 
replaces the original fitting problem\,---\,fit an $N$-term function 
$G_{ab}(t;N)$ to $G_{ab}^\mathrm{MC}(t)$\,---\,by a simpler problem 
that can have far fewer fit parameters: fit an $n$-term function 
$G_{ab}(t;n)$ to ${\tilde G}_{ab}^{\mathrm{MC}}(t;n)$, where $n\!<\!N$. 
Remarkably, as we showed in~\cite{McNeile:2010ji}, \emph{these two 
problems are equivalent for high statistics data even when $n$~is quite 
small:} that is, fit results (means and standard deviations) for the 
low-$j$ parameters are the same in both cases. In the second case, the 
$j\!>\!n$~terms have been ``marginalized,'' or, in effect, integrated 
out of the Bayesian probability distribution, but in a way that does 
not affect the analysis of the $j\!\le\!n$~terms. When $n\!\ll\!N$, the 
fit parameters that remain are many fewer than what would be required 
in a standard fit, and fitting is much faster.

In this paper we use a variation of this marginalization procedure 
which we find to be more robust when fitting correlators that fall 
exponentially quickly with increasing~$t$. In this variation the 
modified correlators are
given by (second definition)
\begin{equation}\label{eq-red}
   \tilde{G}_{ab}^{\mathrm{MC}}(t;n) \equiv
   G_{ab}^\mathrm{MC}(t) \,
   \frac{G_{ab}^\mathrm{pr}(t;n)}{G_{ab}^\mathrm{pr}(t;N)},
\end{equation}
which is analogous to the first definition (\eq{eq-altred}) but applied 
to the logarithm of the correlator rather than the correlator itself. 
Again, terms with $j\!>\!n$ have been removed, and therefore the 
modified correlator data can be fit with the simpler fit 
function,~$G_{ab}(t;n)$.

\begin{figure}
\begin{center}
\includegraphics[scale=0.75]{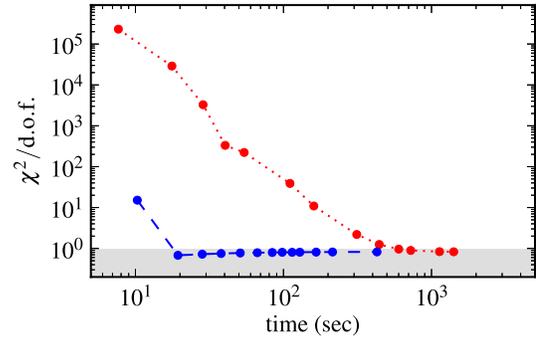}
\vspace{-0.25in}
\end{center}
\caption{\label{fig-upschi2}
Fit $\chi^2$ per degree of freedom for sequential fits of 25~$\Upsilon$ 
correlators with $n\!=\!1,2,3\ldots$ terms in fit 
function~(\ref{eq-fitfcn}). Results are plotted versus the cumulative 
time required for fitting, and are for fits of: a) the unmodified 
simulation data~$G_{ab}^\mathrm{MC}(t)$ (red circles and dotted line); 
and b) the modified simulation data~$\tilde G_{ab}^\mathrm{MC}(t;n)$ 
(\eq{eq-red}) (blue circles and dashed line). The region of good fits 
is indicated by the gray band.
}
\end{figure}

We now illustrate our new method by applying it to QCD simulation data 
from two recent analyses. For each analysis, we fit a function, 
like~$G_{ab}(t;n)$, with $n$~terms both to untouched simulation data 
$G_{ab}^\mathrm{MC}(t)$, and to modified simulation data $\tilde 
G_{ab}^{\mathrm{MC}}(t;n)$, from which $j\!>\!n$~terms have been 
removed using~\eq{eq-red}. We vary~$n$, doing sequential fits with 
$n\!=\!1,2,3\ldots$, where the best-fit parameter values from one fit 
are used as starting values for the next fit. Sequential fitting with 
increasing~$n$ is a standard approach to complicated multi-parameter 
correlator fits; $n$~is increased until the fit's~$\chi^2$ stops 
changing, at which point enough terms have been include to reflect 
accurately the uncertainties introduced by large-$j$ terms. Here we 
examine the best-fit parameters for each~$n$ to investigate the rate at 
which correct results emerge from this process. This allows a detailed 
comparison of our two fitting strategies.

\begin{figure}
\begin{center}
\includegraphics[scale=0.75]{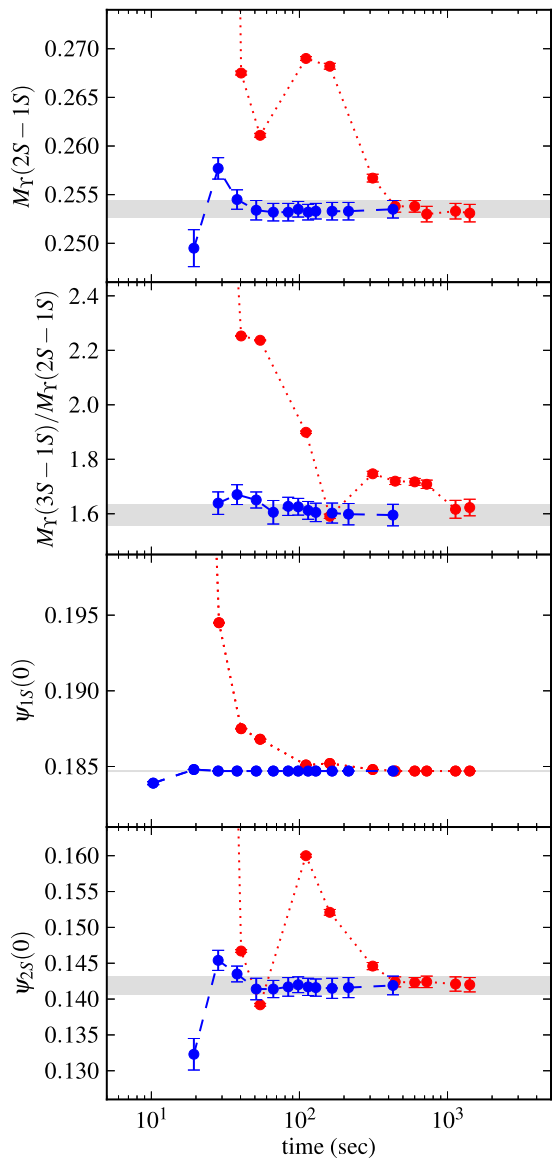}
\vspace{-0.25in}
\end{center}
\caption{\label{fig-ups}
Best-fit results from sequential fits of 25~$\Upsilon$ correlators with 
$n\!=\!1,2,3\ldots$ terms in fit function~(\ref{eq-fitfcn}). Results 
are plotted versus the cumulative time required for fitting, and are 
for fits of: a) the unmodified simulation data~$G_{ab}^\mathrm{MC}(t)$ 
(red circles and dotted line); and b) the modified simulation 
data~$\tilde G_{ab}^\mathrm{MC}(t;n)$ (\eq{eq-red}) (blue circles and 
dashed line). Results are given for mass splittings between different 
vector $S$-states, and for the wave functions at the origin for the 
lowest two states. All results are in lattice units. The gray bands 
show the best-fit result from the modified data after convergence.
}
\end{figure}

The first data set is a collection of 25~correlators for the 
$\Upsilon(1S)$~meson and its radial excitations ($\Upsilon(2S)$, 
$\Upsilon(3S)$, \emph{etc.})~\cite{ups-paper}. These correlators were 
made using five different operators for both sources and sinks. They 
were fit to formula~(\ref{eq-fitfcn}) with priors (in lattice units):
\begin{align}
   \log(E_{1}) &= \log(0.3\pm0.1) = -1.2\pm0.3 \nonumber \\
   \log(E_{j+1}-E_{j}) &= \log(0.25\pm0.125) = -1.4\pm0.5 \nonumber \\
   a_j &= 0.1\pm1.0
\end{align}
except for a local source for which the priors were 
$\log(a_j)\!=\!\log(0.1\pm0.2)\!=\!-2.3\pm2$ (local source). These are 
broad priors\,---\,more than 100~times broader than the final errors 
for the quantities we examine below. We set $N\!=20$ when defining 
$\tilde G_{ab}^\mathrm{MC}(t;n)$~(\eq{eq-red}); this is roughly twice 
the size it needs to be, but it costs little to make $N$~large. In 
general $N$~should be chosen so that terms with $j\!>\!N$ are 
negligible compared with statistical errors.

\begin{figure}
\begin{center}
\includegraphics[scale=0.75]{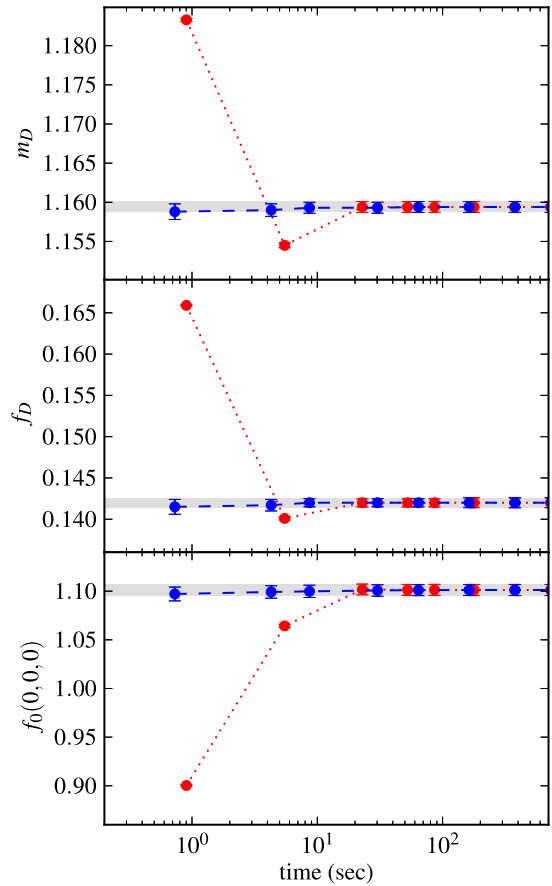}
\vspace{-0.25in}
\end{center}
\caption{\label{fig-D}
Best-fit results from sequential fits of 13~two-point and three-point 
correlators for $D$~and $\pi$~mesons with $n\!=\!1,2,3\ldots$ terms in 
fit function~(\ref{eq-fitfcn}). Results are plotted versus the 
cumulative time required for fitting, and are for fits of: a) the 
unmodified simulation data (red circles and dotted line); and b) the 
modified simulation data~(\eq{eq-red}) (blue circles and dashed line). 
Results are given for the $D$-meson mass~$m_D$ and decay 
constant~$f_D$, and for the $D\!\to\!\pi$ scalar form factor at zero 
recoil momentum~$f_0(0,0,0)$. All results are in lattice units. The 
gray bands show the best-fit result from the modified data after 
convergence.
}
\end{figure}

In Fig.~\ref{fig-upschi2} we plot the $\chi^2$ per degree of freedom 
for each method versus the time required to get to that 
value~\cite{ref-time}. As expected, the new algorithm reaches a 
reasonable $\chi^2$ with just a few terms ($n\!=\!2$--3), in 
20--30~seconds; the traditional algorithm requires $n\!=\!10$--11 to 
obtain a good~$\chi^2$, and 600--700~seconds. Similar differences are 
evident if we look at physical quantities extracted from the 
simulations. In Fig.~\ref{fig-ups} we show results for the $2S-1S$~mass 
splitting (in lattice units), for the $3S-1S$ mass splitting divided by 
the $2S-1S$ splitting, and for the $1S$~and $2S$~mesons' 
(nonrelativistic) wave functions at the origin, which come from fit 
parameters~$a_j$ for a local source. In every case the two algorithms 
agree on the final result, but the new algorithm converges to correct 
results 10--40~times faster.

Our second example is from a recent analysis of the 
$D\!\to\!\pi$~semileptonic form factor~\cite{Na:2011mc}. To extract the 
form factor at four different momenta, this analysis uses a 
simultaneous fit of 13~two-point and three-point correlators: a) a 
$D$-meson correlator with a pseudoscalar local source and sink; b) four 
$\pi$-meson correlators, one for each pion momentum of interest, again 
with local pseudoscalar sources and sinks; and c) two three-point 
correlators $D\!\to\!J_\mathrm{scalar}\!\to\!\pi$ for each of the four 
pion momenta. The fit functions are more complicated for this case. For 
example, the $D$-meson correlator is fit by a function:
\begin{equation} \label{eq-GD}
   G_D(t;n) = \sum_{j=1}^n a_j \,f(E_j,t) - (-1)^{t} a_j^{o}\, 
f(E^o_j,t)
\end{equation}
where $f(E_j,t)\!\equiv\!\exp(-E_jt)+\exp(-E_j(T-t))$ is periodic with 
period $T\!=\!64$, and the second (oscillating) term is due to 
opposite-parity states in the correlator (a feature of staggered-quark 
formalisms like that used in this analysis). The details for the other 
correlators, and the priors are given in~\cite{Na:2011mc}.

\begin{figure}
\begin{center}
\includegraphics[scale=0.8]{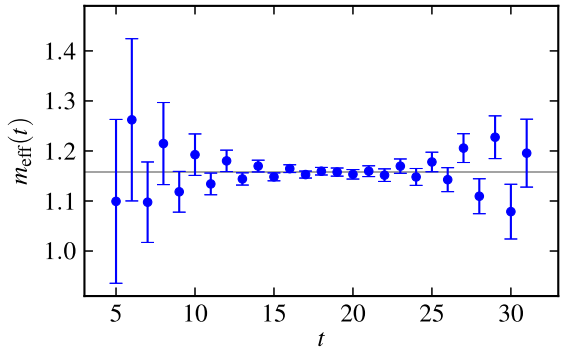}
\vspace{-0.25in}
\end{center}
\caption{\label{fig-meffD}
The $D$-meson's effective mass~$m_\mathrm{eff}(t)$ versus~$t$ computed 
from modified simulation data~$\tilde G_D^\mathrm{MC}(t)$ from which 
every state other than the ground state has been removed (using 
priors). The (very thin) gray band shows the weighted average of all 
$m_\mathrm{eff}(t)$s, taking account of correlations. The thickness of 
the band indicates the uncertainty of the average. Note that the 
largest $t$s shown here correspond to the middle~$t$ range. The error 
bars grow there because $m_\mathrm{eff}(t)$~becomes very sensitive to 
statistical errors in this region (since periodic boundary conditions 
imply that the derivative of the correlator's non-oscillating part 
vanishes at the midpoint).
}
\end{figure}

Despite the complexity of dealing with both two-point and three-point 
correlators, this is a simpler fit than the $\Upsilon$~case; but even 
here we find that marginalizing most of the fit function makes the 
analysis about 30~times faster. We show results in Fig.~\ref{fig-D} for 
the $D$-meson's mass $m_D$ and leptonic decay constant $f_D$, as well 
as for the $D\!\to\!\pi$ scalar form factor~$f_0(0,0,0)$ at zero recoil 
momentum. All results are in lattice units. Again the two approaches 
agree on the results but the new approach has correct results even with 
only a single term ($n\!=\!1$) in the fit functions. For these fits we 
set $N\!=\!10$ when computing the modified data $\tilde 
G_{ab}^\mathrm{MC}(t;n)$~(\eq{eq-red}), which is twice as large as it 
needs to be.

Some insight into how marginalization works can be gained by focusing 
just on the $D$~correlator from this analysis and fitting the modified 
data,
\begin{equation} \label{eq-D}
   \tilde G_D^\mathrm{MC}(t) \equiv G_D^\mathrm{MC}(t)\,
   \frac{a_{1}^{\mathrm{pr}} 
f(E_{1}^\mathrm{pr},t)}{G_D^\mathrm{pr}(t;N)},
\end{equation}
with only the non-oscillating part of the first term 
in~\eq{eq-GD}\,---\,that is, with~$a_1f(E_1,t)$. This situation is 
sufficiently simple that fitting is not required. The $D$~mass, for 
example, can be obtained by averaging the ``effective mass,''
\begin{equation}
   m_\mathrm{eff}(t) \equiv \mathrm{arccosh}\left(
   \frac{\tilde G_D^\mathrm{MC}(t+1)+\tilde G_D^\mathrm{MC}(t-1)}{2\,
   \tilde G_D^\mathrm{MC}(t)}
   \right),
\end{equation}
over all~$t$, taking account of correlations between different~$t$s. 
The effective mass is plotted as a function of~$t$ in 
Fig.~\ref{fig-meffD}. It is compared with the weighted average of 
all~27 $m_\mathrm{eff}(t)$s (gray band), which 
at~$m_\mathrm{eff}^\mathrm{avg}\!=\!1.1584(11)$ agrees well with the 
best result,~$1.1593(7)$, from full multi-term fits (top panel in 
Fig.~\ref{fig-D}).

The first excited state in the $D$~correlator is the opposite-parity 
contribution, which accounts for the oscillation 
in~$m_\mathrm{eff}(t)$. Strong statistical correlations between 
different points result in an average~$m_\mathrm{eff}$ whose error is 
more than 7~times smaller than the best error from an 
individual~$m_\mathrm{eff}(t)$. The errors in~$m_\mathrm{eff}(t)$ when 
$t\!\le\!16$ come almost entirely from marginalized terms absorbed into 
the fit data using~\eq{eq-D}; the original Monte Carlo simulation 
errors are negligible there.

Absent marginalization, contributions from excited states would limit a 
traditional effective mass analysis of this data to values with 
$t\!>\!16$. With marginalization, all $t$s are used, except for a small 
number at very small~$t$ where the fit function is invalid (because of 
temporal non-locality in the lattice quark action). Using 28~$t$s is 
possible because we have removed the excited states through~\eq{eq-D}. 
As a result different $m_\mathrm{eff}(t)$s agree with each other to 
within their errors: fitting all~27 values in Fig.~\ref{fig-meffD} to a 
constant gives an excellent fit, with a $\chi^2$~per degree of freedom 
of~0.6. (The result of the fit is, by definition, the same as the 
weighted average reported above.)

Our new implementation of effective-mass analyses is simpler and less 
ambiguous than traditional analyses because we are not limited to 
large~$t$s. More importantly our implementation also allows us to 
quantify the contribution to the uncertainty in the final 
$m_\mathrm{eff}^\mathrm{avg}$ due to the excited states: here the 
priors for non-oscillating terms in~\eq{eq-GD} contribute 
$0.44\sigma_m$, those from oscillating terms contribute $0.07\sigma_m$, 
and the uncertainties in the Monte Carlo data contribute 
$0.89\sigma_m$, where $\sigma_m$ is the standard deviation of 
$m_\mathrm{eff}^\mathrm{avg}$. Such information is essential for 
assessing the reliability of the final result, as well as for planning 
improvements to the analysis.

In this paper we have shown how to accelerate multi-exponential fits to 
multiple hadronic correlators by removing contributions due to excited 
states from both the fit function and the simulation data, before 
fitting. This technique for marginalizing large parts of the fit 
function greatly reduces the number of fit parameters needed in the 
realistic examples presented here, and makes fitting 10--40~times 
faster. Marginalization also simplifies effective-mass analyses, and 
generalizes easily to analogous multi-state (generalized eigenvalue) 
methods.

This work was supported by the DOE (DE-FG02-04ER41299, 
DE-FG02-91ER40690), the NSF (PHY-0757868), and the STFC. We used the 
Darwin Supercomputer of the Cambridge High Performance Computing 
Service as part of the DiRAC facility jointly funded by STFC, BIS and 
the Universities of Cambridge and Glasgow.  We also used facilities of 
the USQCD collaboration funded by the Office of Science of the DOE and 
at the Ohio Supercomputer Center.

\bibliographystyle{plain}

\begin{thebibliography}{16}
\bibitem{bib-sumexp} Lattice QCD simulations use Euclidean time and 
so~$-it$ is replaced by~$-t$ in the exponentials. Also simulations are 
for finite volumes in space, and therefore all states, including 
multi-hadron states, have discrete energy eigenvalues.
\bibitem{bib-errors} Again the covariance matrix for $\Delta 
G_{ab}(t;n)$ is computed using standard error propagation\,---\,for 
example,
$f(\overline x\pm\sigma_x)\!=\!\overline f\pm\sigma_f$ with $\overline 
f\!\approx\!f(\overline x)$ and 
$\sigma^2_f\!\approx\!f^\prime(\overline x)^2\sigma_x^2$. We have 
compared this linearized analysis with Monte Carlo evaluations of 
$\Delta G$ (from normal distributions for the priors). We find the 
Monte Carlo results to be both much more expensive and also less robust 
for correlators that decay exponentially quickly. Note also that it is 
essential to retain the off-diagonal elements (correlations) in the 
covariance matrix for $\Delta G_{ab}(t;n)$; correlations arise because, 
for example, the prior data used for a parameter is the same for all 
$t$~values.
\bibitem{ref-time} The absolute computer times quoted here are 
obviously of little relevance since they depend upon specific details 
of hardware and software. What is relevant is the comparison between 
methods.
\end{thebibliography}

\end{document}